\documentclass[journal,english,twocolumn]{IEEEtran}
\usepackage[T1]{fontenc}
\usepackage[latin9]{inputenc}
\usepackage{babel}
\usepackage{epsfig}
\usepackage{latexsym}
\usepackage{color}
\usepackage{epsf}
\usepackage{subfig}
\usepackage{amsthm}
\usepackage{amsmath}
\usepackage{graphicx}
\usepackage{amssymb}
\usepackage{psfrag}
\theoremstyle{plain}
 \theoremstyle{plain}
\newtheorem{thm}{Theorem}
 \theoremstyle{definition}
  \newtheorem{example}[thm]{Example}
  \theoremstyle{ }
  \newtheorem{defn}[thm]{Definition}
  \theoremstyle{plain}
  \newtheorem*{cor*}{Corollary}
  \theoremstyle{plain}
  
\newcommand{\argmax}{\operatornamewithlimits{argmax}}
\begin{document}
\title{Energy Efficient Greedy Link Scheduling and Power Control in wireless networks}
\date{}
\author{Arun ~Sridharan, and~ C. Emre ~Koksal\\
Department of Electrical and Computer Engineering\\
The Ohio State University\\
Columbus, Ohio 43210\\
\{sridhara,koksal\}@ece.osu.edu}

\maketitle

\begin{abstract}
We consider the problem of joint link scheduling and power control for wireless networks with average transmission power constraints.  Due to the high
 computational complexity of the optimal policies, we extend the class of greedy link scheduling policies to handle average power constraints. We develop a greedy link scheduling and power control scheme GECS, with provable performance guarantees.
  We show that the performance of our greedy scheduler can
  be characterized using the Local Pooling Factor (LPF) of  a network graph, which has been
   previously used to characterize the stability of the Greedy Maximal Scheduling (GMS) policy for wireless networks. We also simulate the performance of GECS on wireless network, and compare its performance to another candidate greedy link scheduling and power control policy.
\end{abstract}

\vspace{-0.2in}
\section{Introduction}

Energy efficient networking principles have received significant attention in the recent past. Network devices are often provisioned with multiple transmit power levels, but are limited by peak as well as average energy constraints. These constraints are determined by factors such as amplifier characteristics, energy costs and battery limitations. Consequently, one needs efficient control strategies aimed at achieving different performance objectives such as maximizing network lifetime, or maximizing  network throughput while satisfying the energy constraints for the network. Conversely, in  many scenarios, one wants to guarantee a certain network throughput region, while minimizing long term average energy expended to sustain any admissible arrival rate. In this paper, we consider the problem of link scheduling over a wireless network in the presence of average energy constraints.
Owing to the high computational complexity of the optimal policies, we propose low complexity energy efficient link scheduling and power control policies. We exploit the property of Longest Queue First (LQF)-based algorithms to develop a low-complexity greedy power control and scheduling scheme (GECS) for wireless networks under binary interference models. We then analyze its performance in wireless networks  and show that their performance can be characterized using the LPF Of a wireless network. We thus show that the LPF can be used to characterize the performance of cross-layer schemes, spanning multiple layers.\par
A link scheduler selects a non-interfering subset of links in a wireless network to schedule at every instance. A throughput optimal link scheduler is one that keeps the queues in the network stable for the largest set of admissible arrival rates for that network. However, when constraints on average transmission power are imposed, a link scheduler could end up expending more average power than necessary, even when the arrival rate is admissible, as seen from the following example: Consider a single link in which one can either transmit 2 packets in a time slot by expending 2 units of power, or 1 packet in a time slot by expending 0.75 units of power. Consider an arrival process where two packets arrive at the start of every other time slot. The link scheduler would cause the link to transmit whenever there are packets in the queue, \emph{i.e.,} every other time slot, consequently requiring an average power of 2 units. If the average power constraint was less than one units, the link scheduler would be unable to meet the power constraint, whereas by transmitting one packet in each time slot one would consume a lesser average power of 0.75 units and yet guarantee the same throughput.
The same limitations of the link scheduler carries over to the scenario in which multiple interfering links are present. Thus, link scheduling needs to be combined with an effective power control strategy in order to ensure that the largest set of admissible arrival rates can be stabilized while meeting average energy requirements. The optimal link scheduling and power control policy, proposed in \cite{Neely} however, suffers from high computation complexity, which is also one of the main challenges in optimal link scheduling. It is thus natural to ask if one can design low-complexity link scheduling and power control policies, motivated by the study of low-complexity link schedulers. In this context, our contribution is that we extend the scope of greedy schedulers such as Longest-Queue-First and Greedy Maximal Scheduling \cite{LinShroff} to handle energy constraints in wireless networks. Our link scheduling and power control policy, GECS, is based on properties of both greedy schedulers as well as optimal schedulers. Interestingly, we are able to characterize the performance of GECS in terms of the LPF of a network graph, a parameter that has been used to characterize the performance of GMS in wireless networks.

Previous studies have focused on energy efficiency in a particular aspect of network control, (e.g., Energy efficient routing, or energy efficient MAC protocols)  as well as over control decisions spread across the network layers; For instance, \cite{Cruz} and \cite{Kodialam} consider the problem of joint routing, scheduling and power control in wireless networks.
 In \cite{Cruz}, the authors develop a power control and link scheduling policy that minimizes total average power, while satisfying minimum rate constraints over each link. The optimal policy has high computational complexity, exponential in general. The model moreover does not consider  the stochasticity of arrivals and dynamic queue control. In \cite{Kodialam}, the authors provide a low complexity approximation algorithm to solve the joint routing, scheduling and power control problem in wireless networks under a node-exclusive interference model. \par
While the above network optimization problems assume that the long term average rate constraints are known, this is not true in general. Stochastic network optimization considers the problem of dynamic queue control in the presence of unknown arrival rates in order to optimize some quantity such as total average expended power. Stochastic network optimization problems have been addressed using stochastic gradient techniques (\cite{Stolyar, Majumdar}) or Lyapunov optimization techniques \cite{Neely}. In \cite{Neely}, a Lyapunov drift based optimization technique was used to obtain asymptotically optimal policies that minimized the total average power for any arrival rate feasible under the peak power constraints. Lyapunov based optimization  has also been used in energy constrained scheduling, where the objective is to maximize the network throughput region (set of stabilizable arrival rates) subject to average energy constraints. Subsequently, such techniques have been used in dynamic channel acquisition and scheduling to minimize energy (\cite{Neely}, \cite{LiNeely}) consumed for transmission as well as channel acquisition, or to maximize throughput when only limited channel information may be acquired by a link. In all these studies however, the optimal policy has high computational complexity motivating the study of low complexity algorithms with good performance guarantees. While low complexity approximation algorithms have been proposed in specific interference models \cite{Longbi}, or for specific networks\cite{Majumdar}, finding a low complexity dynamic control algorithm with good performance guarantees remains largely unaddressed. \par
In the context of link scheduling, the LQF policy, and its variant Greedy Maximal Scheduling (GMS) have been analyzed extensively as  low-complexity scheduling policy for wireless networks. The performance of GMS has been characterized recently using the Local Pooling Factor, a graph theoretic parameter, and LPF has been studied for a large class of graphs and interference models \cite{LinShroff,Birand}. It is thus beneficial to integrate power control with existing greedy link scheduling policies.
In the next section, we describe the system model. We then describe and analyze the GECS policy in Sections II and III respectively.\vspace{-0.1in}
\section{System Model}\label{sec:model}
We consider a wireless network represented by a network graph $\mathcal{G}=(\mathcal{V},\mathcal{E})$. For simplicity, we initially consider wireless networks with single-hop traffic, so that each link $l$ is a source-destination pair. We assume that link $l$ is associated with interference set $I_l$, so that if link $l$ is active, then no other link in $I_l$ can be active. Every link $l$ is associated with a peak transmit power constraint. We also assume that a transmitter can transmit using a discrete set of power levels $\{P_l^{1},P_l^{2}\cdots P_l^{max}\}$. These discrete power levels are chosen from a convex rate-power curve represented by a function $f_l(c_l)$, where $f_l(c_l)$ is the power required to transmit at rate $c_l$. The function $f_l(c_l)$ depends on the channel conditions of link $l$, which are assumed to be static for the wireless network. For example, in the case of an AWGN channel with a bandwidth $W$, noise spectral density  $N_0$, and a power gain $h_l$,, the power consumed for a rate $c_l$ is given by $\frac{N_0W}{h_l}\left(2^{\frac{c_l}{W}-1}\right)$ (Assuming Gaussian codewords with average power $P_l$). The rates corresponding to the available transmit power levels are denoted by $\{c_l^{1},c_l^{2}\cdots c_l^{max}\}$. \par
We further assume that each link is also associated with an average power constraint $P^{av}_{l}$, so that $\limsup_{T\rightarrow\infty} \frac{1}{T} \sum_{t=1}^{T}P_l(t) \leq P_l^{av}$. We assume that time is slotted so that $P_l(t)$ is the power consumed during time slot $t$. Let $Q_l$ denote the queue of link $l$ into which packets arrive at the beginning of every time slot. Packets arrive into queue $Q_l$ according to an I.I.D. process with finite mean $\lambda_l$.

For a given transmission vector $\vec{P}(t)$, let $\vec{r},\{r_l=f_l^{-1}(P_l)\}$ denote the rate allocation vector achieved by using $\vec{P}(t)$ in time slot $t$. We define a rate allocation vector $\vec{r}$ as being feasible if it satisfies interference constraints, \emph{i.e.,} $r(l)\neq 0 \Rightarrow r(k) =0, \forall k \in I_l,\, k \neq l$. Let $\mathsf{R}=\{\vec{r}\,|\,\vec{r} \text{ is feasible}\}$ and $\mathsf{P}=\{\vec{P}|\,\vec{r} \text{ defined by }r_l=f_l^{-1}(P_l) \text{ is feasible}\}$ denote the set of all feasible rate allocation vectors and power allocation vectors for the network. A scheduling policy $\pi:Q\rightarrow \mathsf{P}$ selects a feasible power allocation vector in each time slot $t$.
\par

\begin{figure}
\begin{centering} \psfrag{P1}{$C_{1}^{av}$} \psfrag{P2}{$C_2^{av}$} \psfrag{Lam2}{$\Lambda_{P^{av}}$}\psfrag{Lam1}{$\Lambda$} \psfrag{c12}{$c_{1}^{max}$}
\psfrag{c22}{$c_{2}^{max}$}\psfrag{l1}{$l_{1}$}
\psfrag{l2}{$l_{2}$}
\includegraphics[bb=0bp 40bp 670bp 350bp,scale=0.32]{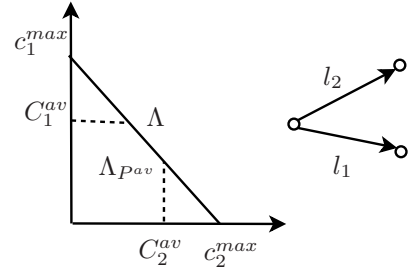}
\par\end{centering}
 \hspace{-0.2in}\caption{Stability region for a two link network graph. $C_{1}^{av}$ and $C_2^{av}$ represent the maximum admissible arrival rate for the two links. The stability region $\Lambda_{P^{av}}$ is the region enclosed within dotted lines. }\label{fig:twolink} \vspace{-0.15in}
\end{figure}

Let $\vec{P}^{av}$ denotes the vector of avaerage power constraints for all links. The power constrained  stability region $\Lambda_{P^{av}}$ can be then defined as \cite{Neely, Tass}:
\begin{align}
&\Lambda_{P^{av}}=\{\vec{\lambda} \mid \exists\, \vec{P}(1),\vec{P}(2),\cdots \text{ such that }\\ \notag
&\lim_{T \rightarrow \infty} \frac{1}{T}\sum_{t=1}^{T} \vec{r}(t) \succ \vec{\lambda} \text{ and } \vec{P}^{av} \succeq \lim_{T \rightarrow \infty} \frac{1}{T}\sum_{t=1}^{T} \vec{P}(t))\},
\end{align}
where $\succ$ and $\succeq$ represent  component wise inequalities.
Fig.~\ref{fig:twolink} shows $\Lambda_{P^{av}}$ for a simple two link network graph with transmit power levels $\mathcal{P}_l=\{0,P_l^{max}\}\,l=1,2$ and average power constraints $P_l^{av}<P_l^{max}$. Links $l_1$ and $l_2$ belong to each other's interference sets. $C_l^{av}$  denotes the highest admissible arrival rate for link $l$. Note that when the available transmit power levels for each link $l$ is increased to $\hat{\mathcal{P}}_l=\{0,P_{l}^1,P_{l}^{max}\}\supset\mathcal{P}_l,\,l=1,2$, then the power constrained stability region increases as well, so that $\Lambda_{P^{av},\mathcal{P}}\subseteq\Lambda_{P^{av},\hat{\mathcal{P}}}$. This follows from the convexity of rate-power curve $f_l(c_l)$. When the average power constraint $P^{av}_l$ is set to $P^{max}_l$, then every link $l$  could always transmit at maximum power $P^{max}_l$ when scheduled, and no power control would be necessary. Otherwise, when $P_l^{max}>P^{av}$, then a joint link scheduling and power control policy is required even when the link channel conditions are static. We are interested in a joint power control and link scheduling policy that can stabilize the queues in the network for any arrival ate vector in $\Lambda_{P^{av}}$ while satisfying the power constraints. \par
In order to satisfy the average power constraints, we employ virtual power queues for each link, as first proposed in \cite{Neely}. Every link $l$ keeps a virtual queue $U_l$ that is used to track the excess power expended beyond the link's power constraint $P_{l}^{av}$. The departures from the virtual queue $U_l$ follow an I.I.D. process with mean $P_{l}^{av}$. The arrivals in each time slot represent the power expended during that slot. Let $P_l(t)$ be the power which link $l$ transmits during time slot $t$, then $P_l$ packets are assumed to have arrived at the beginning of time slot $t+1$. The dynamics of the real and virtual queues in each time slot are then given by:
\[U_l(t+1)=[U_l(t)-P^{av}_{l}(t)]^
+ + P_{l}(t).\] and
\[Q_l(t+1)=[Q_l(t)-S_l(t)]^{+}
+  A_{l}(t).\]
From Lemma 3 of \cite{Neely}, if the virtual queue $U_l$ is stable, then the average power constraint is satisfied. An optimal scheduling policy is one which can stabilize any $\vec{\lambda} \in \Lambda_{P^{av}}$, while ensuring that the power constraints in each link are satisfied.  An optimal scheduling strategy  is given in \cite{Neely} as:
\begin{align}
 \argmax_{P_l, l\in\mathcal{E}} \sum_{l\in\mathcal{E}}(Q_{l}(t)r_l(\vec{P}(t))-U_l(t)P_l(t)) \label{eq:optimal}
\end{align}
The computational complexity of the optimal policy is however, very high, formidably hard with most interference models \cite{GSharma}. Indeed,
even under the binary interference model, eq.~\eqref{eq:optimal} describes a Max-weight policy as the solution maximizes the sum of weights of each link in the selected transmission power vector, where the weight of  link $l$ is given by $ \max_{P_l}Q_lf^{-1}(P_l)-U_lP_l$. The Max-weight scheduler is known to be NP-hard in many cases including K-hop interference models \cite{GSharma}. This motivates the need for a low complexity power control and link scheduling policy with good performance guarantees.\par
In the following section, we describe our link scheduling and power control policy, the Greedy Energy Constrained Scheduler (GECS).  \vspace{-0.12in}
\section{Greedy Energy Constrained Scheduler (GECS)}

%
The GECS scheduler exploits the properties of both LQF as well as Maxweight policies   in order to perform link scheduling and power control. The GECS scheduler first greedily selects links, based on the queue weights $Q_l^2+U_l^2$. It then assigns power levels optimally to the selected link, by applying the optimal power control strategy in \eqref{eq:optimal} as if link $l$ alone is active. This procedure continues until no more links remain to be selected. The GECS scheduler can be formally described as follows:
\begin{itemize}
\item[{(1)}] Select $m \in \argmax_{l \in Z} \{Q_l^2 + U_l^2\}$.
\item[{(2)}] The selected power level satisfies: \newline $P^{*}_{m}=\argmax_{P_m\in\{P_m^{1},P_m^{2}\cdots P_m^{max}\}} Q_m f_{m}^{-1}(P_m)-U_mP_m$
\item[{(3)}] The scheduler then sets $\vec{r}(l)=f_{m}^{-1}(P^{*}_{m})$, and $\vec{r}(k)=0$, for all $k \in I_l\cap Z,\text{ with } k \neq l$.  The set Z is also updated to remove links interfering with $l$ from $Z$.
\item[{(4)}] If $Z \neq \{\emptyset\}$, then repeat step 1.
\end{itemize}

In step 1 as well as step 3, the GECS policy breaks ties randomly. Note that step 2 is only performed on links selected in step 1, and does not increase the computational complexity of the greedy policy, which is $O(|\mathcal{E}|\,log|\mathcal{E}|)$ considering a centralized implementation. The following example illustrates the GECS policy on a 6-cycle network graph under the one-hop interference model.

\begin{figure}
\begin{centering} \psfrag{l1}{$l_1$} \psfrag{l2}{$l_2$} \psfrag{l3}{$l_3$}\psfrag{l4}{$l_4$}\psfrag{l5}{$l_5$}\psfrag{l6}{$l_6$}
\includegraphics[bb=0bp 0 370bp 350bp,scale=0.23]{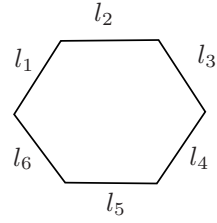}
\par\end{centering}
\caption{\label{fig:six-link-graph}Six-cycle network graph}\vspace{-0.1in}
\end{figure}
\vspace{-0.05in}
\begin{example}
Consider the six cycle graph shown in Fig. \ref{fig:six-link-graph} with links $l_1$ to $l_6$. Let the available transmit power levels for link $l$  be $P_l\in{0,1}$. Let the corresponding rates achieved by $c_l\in\{0,1\}$. Suppose at time $t$, the $Q(t)=[2\; 3\; 8\; 5\; 2\; 10]$, and $U(t)=[1 \;4\; 7\; 5
\; 3\; 12\;]$. Since link $l_6$ has the highest weight $Q_{l}^2+U_{l}^2$,  GECS selects link $l_6$ first. However, in step 3, the transmit power level that maximizes the weight of link $l_6$ is 0. Thus $l_6$ is not scheduled. The next highest weight belongs to $l_3$ and link $l_3$ is added to the schedule. Under the one-hop interference model, links $l_2$ and $l_4$ are interfering links and can not be scheduled. The scheduler then deactivates link $l_5$, and schedules link $l_1$. At the end of this procedure, the rate allocation vector is $\vec{r}=[1\; 0\; 1\; 0\; 0\; 0]$. Note that GECS does not necessarily yield a schedule that is a maximal matching on the graph $\mathcal{G}$.
\end{example}\vspace{-0.05in}

The intuition behind the performance of the GECS scheduler can be explained using  the properties of the LQF and the optimal policies. For any arrival rate vector inside $\Lambda_{P}^{av}$, the optimal policy causes the drift of a quadratic Lyapunov function to be negative. The LQF policy, on the other hand causes the set of longest queues to decrease in the fluid limit if the local pooling conditions are satisfied~\cite{Dimakis}. Since the GECS scheduler combines LQF (for link selection) and the optimal policy (for power control), one expects GECS to cause the drift of a quadratic function to be negative whenever certain local pooling conditions are satisfied. This indeed turns out to be the case, although local pooling occurs in a manner different from \cite{Dimakis}. Nevertheless, the local pooling enables characterizing the performance of GECS using the LPF. In the next section, we analyze the performance of GECS and make these relations precise.
\vspace{-0.15in}
\section{Performance Characterization}

We analyze the performance of the GECS scheduler by characterizing its efficiency ratio. The efficiency ratio of GECS is defined as the largest fraction of $\Lambda_{P}^{av}$, that can be stabilized by GECS. We show that the efficiency ratio of GECS is can be related to the LPF of a network graph, which depends on the topology of the network and the interference model. The LPF can be defined using the local pooling conditions which are given below \cite{JooShroff}:
\begin{defn}
\label{def:def1}Let $L\subset\mathcal{E}$ be any subgraph of $\mathcal{G}$. Let $\vec{s}_L$, where $s_{L}(l)\in\{0,1\}, \,\forall l\in L$  be any $1\times|L|$ link activation vector with the following properties:
(i) $\vec{s}_L$ satisfies interference constraints, and
(ii) $\vec{s}_L$ is maximal, \em{i.e.,} $\exists$ no $l\in L$ with $s_{L}(l)=0$ such that $l$ satisfies interference constraints. Let $\mathcal{M}_L$ be the set of all such link activation vectors $\vec{s}_L$ . Then, $\mathcal{L}$ satisfies $\sigma$-local pooling if, for any given pair  $\vec{\mu},\vec{\nu}$, where $\vec{\mu}$ and $\vec{\nu}$ are convex combinations of elements in $\mathcal{M}_L$, we have $\sigma\vec{\mu}\nprec\vec{\nu}$. \\
The LPF  $\sigma^{*}$, for the network is then defined as:
\begin{align*}
\sigma^{*}=\sup\left\{\sigma\mid\forall\, \mathcal{L}\subset
\hat{\mathcal{G}},
\mathcal{L} \text{ satisfies } \sigma\text{-local pooling} \right\}.
\end{align*}
\end{defn}

In Theorem \ref{Theorem 1}, we show that the efficiency ratio of the $GECS$ policy for a network graph $\mathcal{G}$ is lower bounded by the LPF $\sigma^{*}$ of $\mathcal{G}$.

%


\begin{thm}\label{Theorem 1}
Consider a network graph $\mathcal{G}$ with LPF $\sigma^* $. The $GECS$ policy
stabilizes any $\vec{\lambda}\in\sigma^*\Lambda_{P_{av}}$.
\end{thm}

The proof is given in Appendix A. Although GECS is also in the spirit of the LQF policy, we use a Lyapunov function $V(Q,U)=\max_{l\in\mathcal{E}} Q_{l}^2+U_{l}^2$, that is different from the ones used for LQF and GMS in \cite{Dimakis,LinShroff}. The performance guarantees of GECS in terms of LPF is attractive as the LPF has been extensively studied in context of characterizing the performance of GMS in wireless networks \cite{LinShroff, JooShroff} and \cite{Yexia}. Indeed, \cite{Birand},for instance identifies all network graphs whose LPF is 1 under node-exclusive interference models. Here we show that the LPF can be used to characterize the performance of a joint power controller and scheduler, rather than only the scheduler. This is a step towards a complete performance characterization of cross-layer schemes with a single parameter.\par

\vspace{-0.12in}
\section{Simulations}

\begin{figure}
\noindent \begin{centering} \psfrag{load}{$\lambda_l$}
\includegraphics[bb=50bp 170bp 450bp 600bp,scale=0.32]{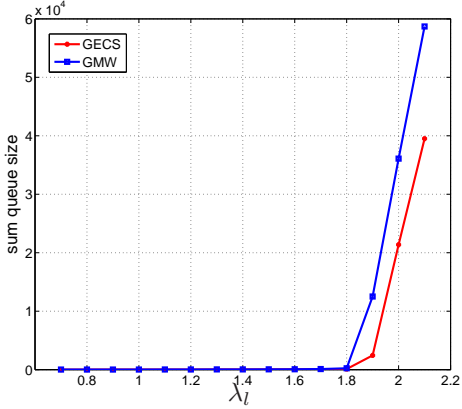}
\par\end{centering}
\centering{}
\noindent \caption{\label{fig:GECScomparison} The performance of GECS and GMW is plotted for the network graph in Fig. \ref{fig:six-link-graph}}
\vspace{-0.25in}
\centering{}
\end{figure}

In this section we simulate the performance of GECS for the six-cycle graph in
Fig.~ \ref{fig:six-link-graph}.
For the same network graph, we also simulate the performance of another candidate greedy scheduler, which we call Greedy Max-weight (GMW). GMW is a greedy approximation of the optimal policy in \eqref{eq:optimal} that gives priority to the link with the highest weight $\argmax_{l\in\mathcal{E}}\{\max_{P_l}(Q_lf^{-1}(P_l)-U_lP_l)\}$. It then proceeds by removing interfering links and repeating this process until no other interfering links remain to be scheduled. Fig.~\ref{fig:GECScomparison} plots the total queue sizes as the offered load $\lambda_l$ in each link is increased towards its maximum admissible arrival rate. The available transmit power levels for each link are assumed to be in the set $\{0,\, 1,\, 3,\,7,\,15\}$ units.  We assume the node exclusive interference model. The average power constraint for each link is set at $2.75$. The corresponding rates are chosen by assuming an AWGN channel so that $C_l=\log(1+P_l\,h_l/N_0)$ where the path loss $h_l$ is arbitrarily chosen for each link but fixed over the scheduling time-frame. The highest admissible arrival rate for each link is given by $[2\; 3.5\; 4.2\; 5.1\; 4.7\; 4]$. Since the LPF of the network graph is 2/3, we expect GECS to simultaneously schedule one third the maximum admissible arrival rate on each link $l$ (since by scheduling 3 of the 6 links at a time, half the maximum admissible arrival rate is admissible for every link)  this is indeed seen to be the case. Although performance guarantees for the GMW policy have not been analyzed using the LPF as we did for GECS in the previous section, the plots show that GECS performs better than GMW at higher loads, and reduces the average sum queue size by 30-75\%. This could be attributed to the fact that GMW only looks at the difference in queue weights and may be insensitive to links that have long queue lengths but with relatively low weight values.
\vspace{-0.15in}
\section{conclusion}
In this paper, we addressed the problem of link scheduling and power control for wireless networks under average power constraints. Owing to the high computational complexity of the optimal policy, we extended the scope of greedy link schedulers to handle average power constraints for wireless networks under a binary interference model. We proposed a greedy link scheduling and power control scheme, GECS, and show that the efficiency ratio of GECS for any wireless network is lower bounded by the Local Pooling Factor (LPF) of the network graph. LPF has been previously used to characterize the performance of certain greedy schedulers, separately from the rest of the system. Here, we show that LPF can be used to characterize the performance of cross-layer schemes, spanning multiple layers.
Some interesting future directions include analyzing the performance of GECS under fading channels, and under more general interference models.
\vspace{-0.17in}
\begin{appendices}

\section{Proof of Theorem \ref{Theorem 1}}
%
\begin{proof}
We prove Theorem 1 by identifying a Lyapunov function and showing that its drift is negative for the fluid limit model of the system. We note that the queuing network can be described using a suitably defined Markov Process , where the states are defined by residual inter-arrival times, residual service times, and queue lengths for all queues in the network. Let $\mathcal{A}_l(t)$ be the process representing the cumulative arrivals upto time $t$,  and $\mathcal{S}_l(t)$ be the cumulative service process for queue $Q_l$. Similarly let $\mathcal{W}_{l}(t)$ and $\mathcal{Y}(t)$ be the cumulative arrival and service process for the virtual queue $U_l(t)$. Consider the sequence of scaled processes $\{\frac{1}{n}\mathcal{A}_l(nt),\frac{1}{n}\mathcal{S}_l(nt),\frac{1}{n}Q_l(nt),\frac{1}{n}\mathcal{W}_{l}(nt), \frac{1}{n}\mathcal{Y}(nt),\frac{1}{n}U_l(nt)\}$. Then, there exists a subsequence $x_n$ such that $\frac{1}{x_n}\mathcal{A}_l(x_nt)\rightarrow\lambda_lt,\frac{1}{x_n}\mathcal{S}_l(x_nt)\rightarrow s_l(t),\frac{1}{x_n}Q_l(x_nt)\rightarrow q_l(t),\frac{1}{x_n}\mathcal{W}_{l}(x_nt)\rightarrow w_l(t), \frac{1}{x_n}\mathcal{Y}_l(x_nt)\rightarrow P^{av}_lt,\frac{1}{x_n}U_l(x_nt)\rightarrow u_l(t)$ w.p. 1. The fluid limits are absolutely continuous, and satisfy:
\begin{align}
&\frac{d}{dt}q_l(t)=
\begin{cases}
[\lambda_l-\pi_l]^{+},& \text{ if } q_l(t)>0 \\
0& \text{ o.w }\\
\end{cases}\\
&\frac{d}{dt}u_l(t)=
\begin{cases}
[\hat{w}_l(t)-P^{av}_l]^{+},& \text{ if } u_l(t)>0 .\\ 
0& \text{ o.w }\\
\end{cases}
\end{align}
Here, $\pi_l(t)=\frac{d}{dt}s_l(t)$ is the service rate into queue $q_l$, whereas
$\hat{w}_l(t)=\frac{d}{dt}w_l(t)$ is the arrival rate into queue $u_l$.
Let $\hat{L}(t)=\argmax_{q_l,u_l}\{q_{l}^2(t)+u_{l}^2(t)\}$.
Let $L(t)=\argmax_{l\in \hat{L}(t)}\frac{d}{dt}(q_{l}^2(t)+u_{l}^2(t))$ be the set of links with the highest derivatives of weights at time $t$. At any regular time $t$, i.e., whenever the derivative exists, we have
$L(t)=\argmax_{l\in \hat{L}(t)}\{q_{l}(t)\left(\lambda-\mu(t)\right)+u_l(t)(\hat{w}_l(t)-P^{av}_l)\}$. Since links in $L(t)$ have the highest derivative of weights, there exists a small $\delta$ such that in the interval $(t,t+\delta)$, weight of links in  $L(t)$ dominates the weight of all other links in the network,\emph{i.e.,} $\min_{l \in L(t)}q_{l}^2(t)+u_{l}^2(t)>\max_{l \in \mathcal{E}/L(t)}q_{l}^2(t)+u_{l}^2(t)$. The GECS policy, therefore gives priority to links in $L(t)$ in the interval $(t,t+\delta)$. We now characterize the service rate of link in the set $L(t)$. We first define the projection of a vector $\vec{a}$ on a set of edges $L$, denoted by $\vec{a}|_{L}$ as $\mathbf{P}\vec{a}$, where $\mathbf{P}$ is a $|L| \times |\vec{a}|_1$ matrix such that $P_{ij}=1,$ if $i=j$ and 0 otherwise. \par
Let $\hat{P}^{*}_l(t)=\argmax_{P_l\in\{P_l^{1},P_l^{2},\ldots,P_l^{max}\}}\{q_l(t)f_l^{-1}(P_l)-u_l(t)P_l\}$, be the set of optimal power levels for link $l \in L(t)$. Let $P^{*}_l(t)=\argmax_{P_l\in \hat{P}_{l}^{*}(t)}\{\frac{d}{dt}q_l(t)f_l^{-1}(P_l)-\frac{d}{dt}u_l(t)P_l\}$. Then, there exists a $\hat{\delta}<\delta$ such that in the time interval $(t,t+\hat{\delta})$, GECS serves link $l\in L(t)$ using power levels in the set $P^{*}_l(t)$.
%
Let $\vec{a}, \{a_{l}\in\{1,0\}\}$ denote a link activation vector that assigns value $1$ to element $a_{l}$ if link $l$ is selected by GECS in step 1 of the algorithm, and 0 if not selected. Since GECS gives priority to links in $L(t)$, the projection of  $\vec{a}$ on $L(t)$ is maximal over $L(t)$ in the interval $(t,t+\hat{\delta})$, \emph{i.e.,} $\vec{a}|_{L(t)}\in\mathcal{M}_{L(t)}$.
To each of the links selected from $L(t)$, GECS assigns power levels from the set $P^{*}_l(t)$. Consequently, any rate allocation vector $\vec{r}$ selected by GECS in the interval $(t,t+\hat{\delta})$, when projected on $L(t)$, satisfies $\vec{r}|_{L(t)}(l)=\{a_l.P_l\}$, where $P_l\in P^{*}_l(t)$ and $\vec{a}\in\mathcal{M}_{L(t)}$. Let $\hat{\mathsf{R}}^{\vec{P}^{*}(t)}_{L(t)}=\{\vec{r}_{L(t)}|\vec{r}_{L(t)}(l)=a_l.P_l, \text{ for some } \vec{a}\in\mathcal{M}_{L(t)}\text{ and } P_l\in {P}_{l}^*(t)\}$ be a set of rate allocation vectors on $L(t)$. The service rate vector $\vec{\mu(t)}$, projected on $L(t)$ can then shown to be a convex combination of elements in $\hat{\mathsf{R}}^{\vec{P}^{*}(t)}_{L(t)}$. A formal argument is omitted here and is similar to the one in \cite{LinShroff}. Any service rate vector $\vec{\mu}(t)$ under the GECS policy can therefore be described using link activation vectors as $\vec{\mu}_l(t)=\sum_j\beta_j\vec{\hat{a}}_j(l)f^{-1}(P_j(l))$, where $P_j(l)\in P^{*}_{l}(t),\,\vec{\hat{a}}_j\in\mathcal{M}_{L(t)}$ and $\sum\beta_j=1$. \par
We then characterize the arrival rate vectors on the set $L(t)$. Let $\mathsf{R}_{L(t)}$ be the set of all feasible rate allocation vectors on the set of link $L(t)$. Since the arrival rate vector $\vec{\lambda} \in \sigma^*\Lambda_{P_{av}}$, its projection on $L(t)$ satisfies $\vec{\lambda}|_{L(t)}\leq \sigma^*\sum_{i}\alpha_i\vec{r}_i$, where $\vec{r}_i\in\mathsf{R}_{L(t)}$ 
Note that any $\vec{r}_{i}$ can be described as $\vec{r}_{i}(l)=a_l.p_l$ for some link activation vector $\vec{a}\in\mathcal{M}_{L(t)}$ and transmission power vector $\vec{P}$. Consequently, any arrival rate vector $\lambda_l$ satisfies $\lambda_l\leq\sigma^*\sum_{i}\alpha_i\vec{a}_i(l)f^{-1}(\vec{P}_{i}(l))$,
 with $\sum_{i}\alpha_i=1$.
We now show that there exists at least one link $l\in L(t)$, such that
$\frac{d}{dt}(q_{l}^2(t)+u_{l}^2(t))<0$. Since $\mathcal{G}$ satisfies $\sigma^*$-local pooling, there exists at least one $\hat{l} \in L(t)$ such that $\sum_{i}\alpha_i\vec{a}_i(l)<\sum_{j}\beta_{j}\vec{\hat{a}}_j(l)$. The transmit power level chosen for this link, $P^{*}_{\hat{l}}(t)$ maximizes $q_{\hat{l}}(t)f^{-1}(P_{\hat{l}})-u_{\hat{l}}(t)P_{\hat{l}}$ across all convex combinations
 of transmission power vectors. Consequently, we have $q_{\hat{l}}\sum_i\alpha_i\vec{a}_i(l)f^{-1}(P_{\hat{l}})-u_l\sum_i\alpha_i\vec{a}_i(l)P_{\hat{l}}<q_{\hat{l}}\sum_j\beta_j\vec{\hat{a}}_j(l)f^{-1}(P^{*}_{\hat{l}})-u_{\hat{l}}\sum_j\beta_j\vec{a}_j(\hat{l})P_{\hat{l}}$, implying that
$q_{\hat{l}}\lambda_{\hat{l}}-u_{\hat{l}}P^{av}_{l}<q_{\hat{l}}\mu_{\hat{l}}(t)-u_{\hat{l}}\hat{w}_{\hat{l}}(t)$. It follows that $\frac{d}{dt}(q_{\hat{l}}^2(t)+u_{\hat{l}}^2(t))<0$, and $\frac{d}{dt}(q_{l}^2(t)+u_{l}^2(t))<0,\,\forall l\in L(t)$. \par
Consider the Lyapunov function $V(t)=\max_{l\in \mathcal{E}}{ q_l^2(t) + u_l^2(t)}$. Then, $\frac{d}{dt}V(t)\leq \max_{l\in\hat{L}(t)}\frac{d}{dt}(q_l^2(t) + u_l^2(t))<0$, establishing the negative drift of $V(t)$. Since the system is stable in the fluid limit, the original system is also stable by the result of Theorem 4.1 in \cite{Dai}.
\end{proof}

\end{appendices}

\end{document}